Research Article

# P vs NP problem in the field anthropology

*Michael .A. Popov, Oxford, UK*

Email Michael282.eps@gmail.com

**Abstract**

This is an attempt of a new kind of anthropology, inspired by assumption that P is not NP, and, where anthropological field ( experimental ) study is considered as a search of a new complexity classes, in particularly, complexity class M.

## Introduction

Real markets are computationally complex and they contain the class socially constructible polynomial Turing machines characterizing what can be computed in the world of *Homo economicus* in practice. There are more than 3000 known **NP-**complete problems ( please, see *Note 1* ) now, including such economics-related problem as Traveling Salesman Problem (*TSP*) . *TSP* ( **NP -** complete problem ) is a problem of combinatorial optimization in economics where given a list of a points and their pairwise distances and the task is to find a shortest possible tour that visits each the point exactly once. We may suppose that *TSP* is not the last **NP-** complete problem, inspired by economics, and , we may await also, that new economic anthropology [ **1** ], used empirical and experimental methods in search of a new complexity classes in different cultural spaces ,can say something new about "platonic reality " of **NP** problems and its practical counter-intuitive solutions .

## Approximation

Let us consider a simplified Cox-Ross-Rubinstein model of an idealized financial market of A-type with limited number of shares where $X(1+n)$, $n = 10, 20, 30, 40, ...$ and $100$ . Hence, a general number of the ways of choosing shares by trader is $2^{10} = 1024$, $2^{20} = 1048576$, $2^{30} = 1073741824$, ... $2^{100} = 1267650600228229401496703205376$, when we assume that each share has only two ways of choosing - " to ignore" / " to buy" .

Such remarkable numbers, transforming an idealized market into true mathematical "monster", are an example of what computer scientists call an **NP**-problem, since there really no feasible way to generate winning formula with the help of a computer during usual working hours by trader ( in polynomial time ), if even we are able to introduce some real boundary meanings, connected with business cycle, demand , management and other parameters of current market situation .

Speaking generally, no future civilization could ever hope to build an optimal ( probably, even quantum ) Turing machine capable to find simplest winning strategies in polynomial time for the real markets of such complexity by "perebor"( in Yablonski's terms),or by brute force search .

Hence, in order to find an anthropological approximation , let us to imagine a kind of anthropological field study used *quantum – like game* [ *Note 2* ]**,** where player Bob is a skeptic ,and, he does not believe that such sort of **NP** problem ( formulated for some local supermarket) can be solved by brutal force search, whereas Alice, another player, suggests that **NP** has practical solution **M**, if even **NP** is not **P.**

Thus, Bob issues a challenge to Alice: he choses **NP**-problem for which Alice cannot have known efficient solution Hence, if Alice is able to find some M = **NP** in polynomial time, Bob can easily convince himself that Alice is winner .

Because $P \neq NP$ **a**ssumption can have fundamental meaning for natural and social sciences ; in 2005-2008 I made an attempt to translate this kind of game into a practical game ( called "Oxford game") in order to demonstrate real existence of **M**-decisions.

**Experiment**

In my "*Oxford P vs. NP game*", collective player Bob is skeptical customers of T supermarket which believe that such **NP-** complete problem as exact prediction of the real ways of choosing goods, used by customers in certain day ( or even hour) from certain area of the shop floor, cannot be easy polynomially computed, or, **P ≠ NP**.

Indeed, if we take general number of the offers ( goods) in the given T supermarket ,say, exactly 10 235, then, speaking mathematically, a general number of the ways of choosing goods by customers is truly impossible astronomical number:

$$2^{10235} = 1.1015535662838695803546296374376e+3081$$

( when we assume that each offer has only two ways of choosing - " to ignore" / " to buy" and that number of supermarket's goods is not changing in a given time ).

Such surprisingly astronomical number also suggests that if even we should introduce some real boundary limits, connected with season, culture, shop's strategy to use reduced prices, weather, income and local habits of the customers, we also cannot essentially reduce such number to any reasonable number as well (for instance if we select just only 100 basic goods we'll have astronomical number

$$2^{100} = 1267650600228229401496703205376 \text{ of the ways of choice }).$$

Thus, speaking in the terms of computational complexity theory, this is an example of a **NP -** problem. Moreover, because any successful trader cannot use any conventional or even non-conventional computer in time polynomial in the input size at shop floor**,** we, hence, cannot suggest that **P= NP**, here.

In other words, if player B's assumption is right, it is impossible to imagine an existence of successful trader Alice, having a proof of an existence of non-**P** or **M** solution for **NP**. Hence, like in any two-person zero-sum game, the competitive aspect is extremely expressed here, since whatever is won by Bob is lost by the other player Alice. Thus, Alice wins if A can demonstrate an existence of **M**-solution In other words, there must exist impossible nonlocal **EPR**-like correlation between counter-intuitive activity of A and synchronic trend of sales.

In our game with trader A, we can accept the usual restriction of zero - sum games that each player has only a finite number of strategies. Random moves are permitted, correspondingly, the game might not be deterministic. There is no requirement of perfect information, that is, a player does not necessarily know what moves the opponent -trader A has made.

Assume that player A ( successful trader Alice ) has two strategies – to be at shop floor (**A'**) and to be out shop floor (**A"**) and that player B has two strategies also, namely: to buy goods (**B'**) or to ignore goods (**B"**).

Thus, the game may be described by normal payoff bi-matrix for A in the form :

|  |  | **Bob** | |
|---|---|---|---|
|  |  | strategy **B'** | strategy **B"** |
| **Alice** | strategy **A'** | 1 | 0 |
|  | strategy **A"** | 0 | 1 |

Hence, the matrix of the game in normal form need not have any saddle point at all and the game is inherently unstable.

Our systematic observations of 2007-2008 in Oxford's supermarket T on work of successful trader( "Michael") suggest that counter-intuitively successful trader A can have real **M** decisions in polynomial time. Systematic review of the Oxford gaming may be represented in the following manner.

## Table 1

| Date of game | Strategy of A | Strategy of B | Matrix | Synchronic trend of sales |
|---|---|---|---|---|
| 16.02.2007 | A'' | B'' | (0  1) | down |
| 17.02.2007 | A'' | B'' | (0  1) | down |
| 23.02.2007 | A'' | B'' | (0  1) | down |
| 24.02.2007 | A' | B' | (1  0) | up |
| 16.03.2007 | A'' | B'' | (0  1) | down |
| 17.03.2007 | A'' | B'' | (0  1) | down |
| 30.03.2007 | A' | B'' | (0  1) | down |
| 02.04.2007 | A'' | B'' | (0  1) | down |
| 08.04.2007 | A' | B' | (1  0) | up |
| 14.04.2007 | A'' | B'' | (0  1) | down |
| 16.04.2007 | A'' | B'' | (0  1) | down |
| 20.04.2007 | A'' | B'' | (0  1) | down |
| 21.04.2007 | A' | B' | (1  0) | up |
| 23.04.2007 | A' | B' | (1  0) | up |
| 27.04.2007 | A'' | B'' | (0  1) | down |
| 28.04.2007 | A' | B' | (1  0) | up |
| 30.04.2007 | A'' | B'' | (0  1) | down |
| 4.05.2007 | A' | B' | (1  0) | up |
| 5.05.2007 | A' | B' | (1  0) | up |
| 11.05.2007 | A'' | B'' | (0  1) | down |
| 12.05.2007 | A' | B' | (1  0) | up |
| 14.05.2007[7.30-10.00] | A' | B' | (1  0) | up |
| 14.05.2007[10.00-3.30] | A'' | B'' | (0  1) | down |
| 18.05.2007[6.00-9.00] | A' | B' | (1  0) | up |
| 18.05.2007[9.00-1.00] | A'' | B'' | (0  1) | down |
| 19.05.2007 | A' | B' | (1  0) | up |
| 21.05.2007[7.30-10.00] | A' | B' | (1  0) | up |
| 21.05.2007[10.00-3.30] | A'' | B'' | (0  1) | down |

| Date | | | | |
|---|---|---|---|---|
| 25.05.2007 | A' | B' | (1  0) | up |
| 26.05.2007 | A' | B' | (1  0) | up |
| 1.06.2007 | A'' | B'' | (0  1) | down |
| 2.06.2007[7.30-10.00] | A' | B' | (1  0) | up |
| 2.06.2007[10.00-3.30] | A'' | B'' | (0  1) | down |
| 4.06.2007 | A' | B' | (1  0) | up |
| 8.06.2007 | A' | B' | (1  0) | up |
| 9.06.2007 | A' | B' | (1  0) | up |
| 11.06.2007 | A' | B' | (1  0) | up |
| 15.06.2007[9.00-11.45] | A'' | B'' | (0  1) | down |
| 15.06.2007[12.00-1.30] | A' | B' | (1  0) | up |
| 16.06.2007 | A' | B' | (1  0) | up |
| 18.06.2007[7.35-2.30] | A'' | B'' | (0  1) | down |
| 18.06.2007[2.30-3.30] | A' | B' | (1  0) | up |
| 22.06.2007 | A'' | B'' | (0  1) | down |
| 23.06.2007 | A' | B' | (1  0) | up |
| 29.06.2007 | A'' | B'' | (0  1) | down |
| 30.06.2007 | A'' | B'' | (0  1) | down |
| 2.07.2007[7.30-10.00] | A' | B' | (1  0) | up |
| 2.07.2007[10.15-3.30] | A'' | B'' | (0  1) | down |
| 6.07.2007[6.00-9.00] | A' | B' | (1  0) | up |
| 6.07.2007[9.30-1.00] | A'' | B'' | (0  1) | down |
| 7.07.2007 | A' | B' | (1  0) | up |
| 13.07.2007 | A'' | B'' | (0  1) | down |
| 14.07.2007 | A'' | B'' | (0  1) | down |
| 16.07.2007[7.30-10.00] | A' | B' | (1  0) | up |
| 16.07.2007[10.00-3.30] | A'' | B'' | (0  1) | down |
| 20.07.2007 | A' | B' | (1  0) | up |
| 21.07.2007 | A'' | B'' | (0  1) | down |
| 23.07.2007 | A'' | B'' | (0  1) | down |
| 27.07.2007 | A' | B' | (1  0) | up |

| Date | | | | |
|---|---|---|---|---|
| 28.07.2007[730-1.00] | A'' | B'' | (0  1) | down |
| 28.07.2007[1.00-3.30] | A' | B' | (1  0) | up |
| 30.07.2007 | A' | B' | (1  0) | up |
| 3.08.2007 | A' | B' | (1  0) | up |
| 4.08.2007 | A' | B' | (1  0) | up |
| 6.08.2007 | A' | B' | (1  0) | up |
| 10.08.2007 | A' | B' | (1  0) | up |
| 11.08.2007 | A' | B' | (1  0) | up |
| 13.08.2007[7.30-10.00] | A' | B' | (1  0) | up |
| 13.08.2007[10.30-2.30] | A'' | B'' | (0  1) | down |
| 13.08.2007[2.30-3.30] | A' | B' | (1  0) | up |
| 17.08.2007[6.00-9.00] | A' | B' | (1  0) | up |
| 17.08.2007[9.00-1.00] | A'' | B'' | (0  1) | down |
| 18.08.2007 | A' | B' | (1  0) | up |
| 20.08.2007 | A' | B' | (1  0) | up |
| 24.08.2007 | A'' | B'' | (0  1) | down |
| 25.08.2007 | A'' | B'' | (0  1) | down |
| 31.08.2007 | A' | B' | (1  0) | up |
| 1.09.2007[7.30-9.00] | A' | B' | (1  0) | up |
| 1.09.2007[9.30-11.15] | A'' | B'' | (0  1) | down |
| 1.09.2007[11.30-2.00] | A' | B' | (1  0) | up |
| 1.09.2007[2.30-3.30] | A'' | B'' | (0  1) | down |
| 3.09.2007 | A'' | B'' | (0  1) | down |
| 7.09.2007[8.00-10.00] | A'' | B'' | (0  1) | down |
| 7.09.2007[10.30-1.00] | A' | B' | (1  0) | up |
| 8.09.2007 | A' | B' | (1  0) | up |
| 10.09.2007[7.30-10.00] | A' | B' | (1  0) | up |
| 10.09.2007[10.30-3.30] | A'' | B'' | (0  1) | down |
| 14.09.2007 | A'' | B'' | (0  1) | down |
| 15.09.2007[7.30-10.00] | A'' | B'' | (0  1) | down |
| 15.09.2007[10.30-3.30] | A' | B' | (1  0) | up |

| Date | | | | |
|---|---|---|---|---|
| 17.09.2007 | A'' | B'' | (0  1) | down |
| 21.09.2007[6.00-10.00] | A' | B' | (1  0) | up |
| 21.09.2007[10.30-1.00] | A'' | B'' | (0  1) | down |
| 22.09.2007 | A' | B' | (1  0) | up |
| 24.09.2007 | A' | B' | (1  0) | up |
| 28.09.2007[6.00-10.00] | A' | B' | (1  0) | up |
| 28.09.2007[10.00-1.00] | A'' | B'' | (0  1) | down |
| 29.09.2007[2.30-3.30] | A' | B' | (1  0) | up |
| 1.10.2007 | A' | B' | (1  0) | up |
| 5.10.2007[6.00-9.00] | A' | B' | (1  0) | up |
| 5.10.2007[9.00-1.00] | A'' | B'' | (0  1) | down |
| 6.10.2007[7.30—9.00] | A' | B' | (1  0) | up |
| 6.10.2007[9.30-12.15] | A'' | B'' | (0  1) | down |
| 6.10.2007[12.30-3.30] | A' | B' | (1  0) | up |
| 12.10.2007 | A' | B' | (1  0) | up |
| 13.10.2007 | A' | B' | (1  0) | up |
| 15.10.2007[7.30-9.00] | A' | B' | (1  0) | up |
| 15.10.2007[9.15-1.00] | A'' | B'' | (0  1) | down |
| 15.10.2007[2.30-3.30] | A' | B' | (1  0) | up |
| 19.10.2007[6.00-9.00] | A' | B' | (1  0) | up |
| 19.10.2007[9.15-1.00] | A'' | B'' | (0  1) | down |
| 20.10.2007 | A' | B' | (1  0) | up |
| 22.10.2007 | A' | B' | (1  0) | up |
| 26.10.2007 | A' | B' | (1  0) | up |
| 27.10.2007 | A' | B' | (1  0) | up |
| 29.10.2007[7.30-10.00] | A' | B' | (1  0) | up |
| 29.10.2007[10.30-2.00] | A'' | B'' | (0  1) | down |
| 29.10.2007[2.30-3.30] | A' | B' | (1  0) | up |
| 2.11.2007[6.00-9.00] | A' | B' | (1  0) | up |
| 2.11.2007[9.30-11.00] | A'' | B'' | (0  1) | down |
| 2.11.2007[11.00-1.00] | A' | B' | (1  0) | up |

| Date | | | | |
|---|---|---|---|---|
| 3.11.2007[7.30-10.00] | A' | B' | (1  0) | up |
| 3.11.2007[10.30-1.48] | A'' | B'' | (0  1) | down |
| 3.11.2007[2.30-3.30] | A' | B' | (1  0) | up |
| 5.11.2007[7.30-10.00] | A' | B' | (1  0) | up |
| 5.11.2007[10.30-1.50] | A'' | B'' | (0  1) | down |
| 5.11.2007[2.30-3.30] | A' | B' | (1  0) | up |
| 9.11.2007[6.00-10.00] | A'' | B'' | (0  1) | down |
| 9.11.2007[10.30-1.00] | A' | B' | (1  0) | up |
| 10.11.2007[7.30-8.40] | A' | B' | (1  0) | up |
| 10.11.2007[9.15-10.45] | A'' | B'' | (0  1) | down |
| 10.11.2007[10.55-3.30] | A' | B' | (1  0) | up |
| 12.11.2007[8.50—9.30] | A'' | B'' | (0  1) | down |
| 12.11.2007[10.00-3.30] | A' | B' | (1  0) | up |
| 16.11.2007 | A' | B' | (1  0) | up |
| 17.11.2007 | A' | B' | (1  0) | up |
| 23.11.2007 | A' | B' | (1  0) | up |
| 24.11.2007 | A' | B' | (1  0) | up |
| 26.11.2007 | A' | B' | (1  0) | up |
| 1.12.2007 | A' | B' | (1  0) | up |
| 4.12.2007 | A'' | B'' | (0  1) | down |
| 7.12.2007 | A' | B' | (1  0) | up |
| 8.12.2007 (from 11 am) | A' | B' | (1  0) | up |
| 10.12.2007 | A' | B' | (1  0) | up |
| 14.12.2007 | A' | B' | (1  0) | up |
| 15.12.2007 | A' | B' | (1  0) | up |
| 17.12.2007 | A' | B' | (1  0) | up |
| 21.12.2007 | A' | B' | (1  0) | up |
| 22.12.2007 | A' | B' | (1  0) | up |
| 28.12.2007 | A' | B' | (1  0) | up |
| 29.12.2007[7.30-9.00] | A' | B' | (1  0) | up |
| 29.12.2007[9.30-12.00] | A'' | B'' | (0  1) | down |

| Date | | | | |
|---|---|---|---|---|
| 29.12.2007 [2.30-3.30] | A' | B' | (1  0) | up |
| 20.01.2008 | A' | B' | (1  0) | up |
| 22.01.2008 | A' | B' | (1  0) | up |
| 25.01.2008 | A' | B' | (1  0) | up |
| 26.01.2008 | A'' | B'' | (0  1) | down |
| 1.02.2008 | A' | B' | (1  0) | up |
| 2.02.2008 | A'' | B'' | (0  1) | down |
| 8.02.2008 | A'' | B'' | (0  1) | down |
| 9.02.2008 | A'' | B'' | (0  1) | down |
| 11.02.2008 | A'' | B'' | (0  1) | down |
| 15.02.2008 | A'' | B'' | (0  1) | down |
| 16.02.2008 | A'' | B'' | (0  1) | down |
| 18.02.2008 | A'' | B'' | (0  1) | down |
| 22.02.2008 | A' | B' | (1  0) | up |
| 29.02.2008 [from 9 am] | A'' | B'' | (0  1) | down |
| 1.03.2008 | A' | B' | (1  0) | up |
| 3.03.2008 | A' | B' | (1  0) | up |
| 7.03.2008 | A' | B' | (1  0) | up |
| 8.03.2008 | A' | B' | (1  0) | up |
| 10.03.2008 | A' | B' | (1  0) | up |
| 22.03.2008 | A' | B' | (1  0) | up |
| 28.03.2008 | A' | B' | (1  0) | up |
| 29.03.2008 | A' | B' | (1  0) | up |
| 31.03.2008 | A' | B' | (1  0) | up |
| 7.04.2008 | A' | B' | (1  0) | up |
| 11.04.2008 | A' | B' | (1  0) | up |
| 12.04.2008 | A' | B' | (1  0) | up |
| 14.04.2008 | A' | B' | (1  0) | up |
| 18.04.2008 | A' | B' | (1  0) | up |
| 19.04.2008 | A' | B' | (1  0) | up |
| 21.04.2008 | A' | B' | (1  0) | up |
| 25.04.2008 | **A'** | **B'** | (1  0) | up |
| 26.04.2008 | **A'** | **B'** | (1  0) | up |
| 28.04.2008 | **A''** | **B''** | (0  1) | down |
| 2.05.2008 | **A'** | **B'** | (1  0) | up |
| 9.05.2008 | **A'** | **B'** | (1  0) | up |
| 10.05.2008 | **A'** | **B'** | (1  0) | up |
| 12.05.2008 | **A'** | **B'** | (1  0) | up |
| 16.05.2008 | **A'** | **B'** | (1  0) | up |
| 17.05.2008 | **A'** | **B'** | (1  0) | up |
| 19.05.2008 | **A'** | **B'** | (1  0) | up |
| 23.05.2008 | **A'** | **B'** | (1  0) | up |
| 24.05.2008 | **A'** | **B'** | (1  0) | up |
| 30.05.2008 | **A'** | **B'** | (1  0) | up |
| 31.05.2008 | **A'** | **B'** | (1  0) | up |
| 2.06.2008 | **A'** | **B'** | (1  0) | up |
| 6.06.2008 | **A'** | **B'** | (1  0) | Up |

**Conclusion**

In the context of our results we may assume that player B's assumption is not correct, it is really possible to imagine even in complex markets the existence of successful trader Alice, having a counter-intuitive proof of an existence of non-**P** or M solution for **NP**. Correspondingly, there is "impossible" correlation between M-decision practice of **A** and synchronic trend of sales, when the total number of ways of choosing successful winning combination by brute force search in polynomial time( 7 working hours ) is completely impractical.

Market philosophy during last two hundred years was formulated under influence of the classical probability, classical theory of games and belief in existence of some kind of " a Simplicity Island" that could be easy found by pure thinking without market experiments.

Quantum games discovered another opportunities to understand markets in good agreement with some fundamental experiments of quantum physicists of 1990s and "holistic visions" by anthropologists. Thus, we may conclude that complexity economics and complexity anthropology could be considered as a kind of optimistic paradigm for practical understanding of the financial crisis now.

*Note 1*

The **P versus NP** problem is fundamental problem of computer science and computation complexity theory, where it is supposed that there are two classes of natural problems: **P** class solvable by deterministic algorithm in polynomial time and NP class of problems solvable by nondeterministic algorithm in polynomial time. In his 20 March 1956 letter to von Neumann, K. Gõdel developed surprisingly modern formulation for **P =NP** understanding. Accordance with letter, let us suppose that there exists a Turing machine such that for every formula F in first order predicate logic and every natural number which allow us to decide if there is a proof ( demonstration of an existence of the natural number ,for instance in number theory ) of F of length n ( n-number of symbols -digits ). Hence, let $\Phi(F,n)$ be the number of steps the machine requires for this and let $\varphi(n) = max'\ \Phi(F,n)$. Thus, the problem **P = ?NP** could be expressed as the question how fast $\varphi(n)$ grows for some imaginary optimal machine ? If such kind of Turing machine and a natural number d such that $\varphi(n) < n^d$ exist, then **P = NP.** Thus, hence, the **NP**-complete problem could be defined as the case when $\varphi(n) = \infty = max'\ \Phi(F,n)$. Correspondingly, when $M < \varphi(n) < n^d$ exists, we have **M = NP** and **P ≠ NP.**

An understanding **P vs NP** problem is of great practical, philosophical and anthropological significance. In particular, in accordance with K. Gõdel, dramatic equivalentness **P=NP** means that " the mental work of a mathematician concerning Yes-or -No questions could be completely replaced by a machine." [ **2** ]

*Note 2*

Quantum games are non-classical games, based on quantum effect, called Entanglement [ **3** ]. Quantum-like or pseudo-quantum games where two entangled players ( without quantum devices ) can communicate non-locally are merely macroscopic approximation to "true"quantum games in physics. Quantum-like games could be found now in physical,biological& economical literature  [ **4** ] [ **5** ] [ **6** ]